\documentclass[a4paper,fleqn,usenatbib]{mnras} 
\usepackage[T1]{fontenc}
\usepackage{ae,aecompl}

\newcommand{\beq}{\begin{equation}}
\newcommand{\eeq}{\end{equation}}
\newcommand{\beqa}{\begin{equation}\begin{aligned}}
\newcommand{\eeqa}{\end{aligned}\end{equation}}

\newcommand{\Msun}{\rm M_\odot}

\newcommand{\LUV}{L_{\rm UV}}
\newcommand{\LIR}{L_{\rm IR}}
\newcommand{\KUV}{K_{\rm UV}}
\newcommand{\KIR}{K_{\rm IR}}

\newcommand{\IRX}{{\rm IRX}}
\newcommand{\SFR}{{\rm SFR}}

\newcommand{\SSFR}{{\rm SSFR}}

\newcommand{\Ms}{M_{*}}

\newcommand{\vpk}{v_{\rm peak}}

\usepackage{mathtools,tabu}
\usepackage{amsmath,amssymb}

\title[Minimal empirical model for CFIRB]{A minimal empirical model for the cosmic far-infrared background anisotropies}
\author[H.-Y.\ Wu and O.\ Dor\'e]{Hao-Yi Wu$^{1,2}$\thanks{E-mail: hywu@caltech.edu} and Olivier Dor\'e$^{1,2}$\\
$^{1}$California Institute of Technology, 1200 E.\ California Blvd., MC 367-17, Pasadena, CA 91125, USA \\
$^{2}$Jet Propulsion Laboratory, California Institute of Technology, 4800 Oak Grove Drive, Pasadena, CA 91109, USA}
\date{Accepted 2017 January 4. Received 2016 December 30; in original form 2016 November 22}
\pubyear{2017}
\begin{document}
\label{firstpage}
\pagerange{\pageref{firstpage}--\pageref{lastpage}}
\maketitle

\begin{abstract}
Cosmic far-infrared background (CFIRB) probes unresolved dusty star-forming galaxies across cosmic time and is complementary to ultraviolet and optical observations of galaxy evolution. In this work, we interpret the observed CFIRB anisotropies using an empirical model based on resolved galaxies in ultraviolet and optical surveys.  Our model includes stellar mass functions, star-forming main sequence, and dust attenuation. We find that the commonly used linear Kennicutt relation between infrared luminosity and star formation rate overproduces the observed CFIRB amplitudes.  The observed CFIRB requires that low-mass galaxies have lower infrared luminosities than expected from the Kennicutt relation, implying that low-mass galaxies have lower dust content and weaker dust attenuation.  Our results demonstrate that CFIRB not only provides a stringent consistency check for galaxy evolution models but also constrains the dust content of low-mass galaxies.

\end{abstract}


\section{Introduction}

Cosmic far-infrared background (CFIRB) originates from unresolved dusty star-forming galaxies from all redshifts and accounts for half of the extragalactic background light generated by galaxies.  In dusty star-forming galaxies, $\sim90\%$ of the ultraviolet (UV) photons produced by recent star-forming activities are absorbed by interstellar dust and re-emitted in far-infrared (FIR; also known as submillimeter, hereafter submm; 100--1000 $\micron$).   The FIR luminosities of galaxies are thus tracers of star formation rate  (SFR) and are complementary to UV luminosities \citep[e.g.,][]{Kennicutt98,KennicuttEvans12,MadauDickinson14}.  Compared with UV, galaxies are much less understood in FIR/submm due to the low-resolution of telescopes in these wavelengths.  Despite the recent progress in resolving galaxies in FIR/submm \citep[e.g.,][]{Casey14,Lutz14,Dunlop16,Fujimoto16,Geach16}, most of the dusty star-forming galaxies remain unresolved.  Therefore, CFIRB provides a rare opportunity to study dusty star-forming galaxies under the current resolution limit.
 
First predicted by \cite{PartridgePeebles67b} and \cite{Bond86}, CFIRB was discovered by {\em COBE}-FIRAS, which also provided to date the only absolute intensity measurement of CFIRB \citep{Puget96,Fixsen98,Hauser98,Gispert00,HauserDwek01}.  
Thereafter, the anisotropies of CFIRB have been measured to ever-improving accuracy by 
{\em Spitzer} \citep{Lagache07}, 
BLAST \citep{Viero09}, 
SPT \citep{Hall10}, 
{\em AKARI} \citep{Matsuura11}, 
ACT \citep{Hajian12}, 
{\em Herschel}-SPIRE \citep{Amblard11,Berta11,Viero13},  
and {\em Planck}-HFI \citep{Planck11CIB, Planck13XXX}. 
In addition, CFIRB maps have been cross-correlated with the lensing potential observed using cosmic microwave background \citep[CMB,][]{Planck13XVIII} and with near-infrared background \citep[][]{Thacker15}.

The CFIRB anisotropies have been interpreted mostly using phenomenological models \citep[e.g.,][]{Viero09,Amblard11,Planck11CIB,DeBernardis12,Shang12,Xia12, Addison13,Viero13,Planck13XXX}.  Although these models can fit the data, they provide limited insight into the underlying galaxy evolution processes. Since galaxy evolution has been extensively studied by UV/optical surveys, it is necessary to understand whether CFIRB agrees with the current knowledge of galaxy evolution. 

In this work, we construct an empirical model for dusty star-forming galaxies based on recent galaxy survey results, including stellar mass functions, star-forming main sequence, and dust attenuation.   We find that, without introducing new parameters, a minimal model can well reproduce the observed CFIRB anisotropies and submm number counts.  Our model is the first step towards constructing a comprehensive model for UV, optical, and FIR observations, as well as building multiwavelength mock catalogues for these observations.  Such a model is essential for the understanding of cosmic star-formation history and for extracting the most information from multiwavelength surveys.
 
Our approach is similar to the empirical approach adopted by \cite{BetherminDore12} and \cite{BetherminDaddi12,Bethermin13}.  Our major innovations include using an $N$-body simulation and recent self-consistent compilations of stellar mass functions and star-forming main sequence.  We also adopt a minimalist approach; that is, we look for the simplest, observationally-motivated model that agrees with CFIRB observations.  In each step of our modelling, we directly use constraints from recent observations and avoid introducing new parameters or fitting model to the data. This work is complementary to our earlier work of interpreting CFIRB using a physical gas regulator model \citep{Wu16}.

This paper is organized as follows. We introduce our model in Section~\ref{sec:model} and calculate the CFIRB anisotropies in Section~\ref{sec:obs}. Section~\ref{sec:results} compares our model predictions with the observational results of {\em Planck} and {\em Herschel}. We discuss our results in Section~\ref{sec:discussions} and summarize in Section~\ref{sec:summary}. Throughout this work, we use the cosmological parameters adopted by the Bolshoi--Planck simulation (see Section~\ref{sec:BolshoiP}), the stellar population synthesis (SPS) model from \citet[][BC03]{BruzualCharlot03},
and the initial mass function (IMF) from \cite{Kroupa01}. 

\section{Empirical Model}\label{sec:model}

We construct a model to generate the infrared (IR) spectral flux densities $S_\nu$ for a population of galaxies.  Our model includes following five steps:

\begin{enumerate}

\item Sampling dark matter haloes from the Bolshoi--Planck simulation (Section~\ref{sec:BolshoiP})

\item Performing abundance matching to assign stellar mass ($\Ms$) to haloes (Section~\ref{sec:abmatch})

\item Assigning $\SFR$ to $\Ms$ based on the star-forming main sequence (Section~\ref{sec:SFR_Ms})

\item Calculating IR luminosity ($\LIR$) based on SFR and $\Ms$ (Section~\ref{sec:LIR_SFR})

\item Calculating $S_\nu$ by assuming a spectral energy distribution (SED; Section~\ref{sec:SED})

\end{enumerate}
 
\begin{figure*}
\includegraphics[width=0.67\columnwidth]{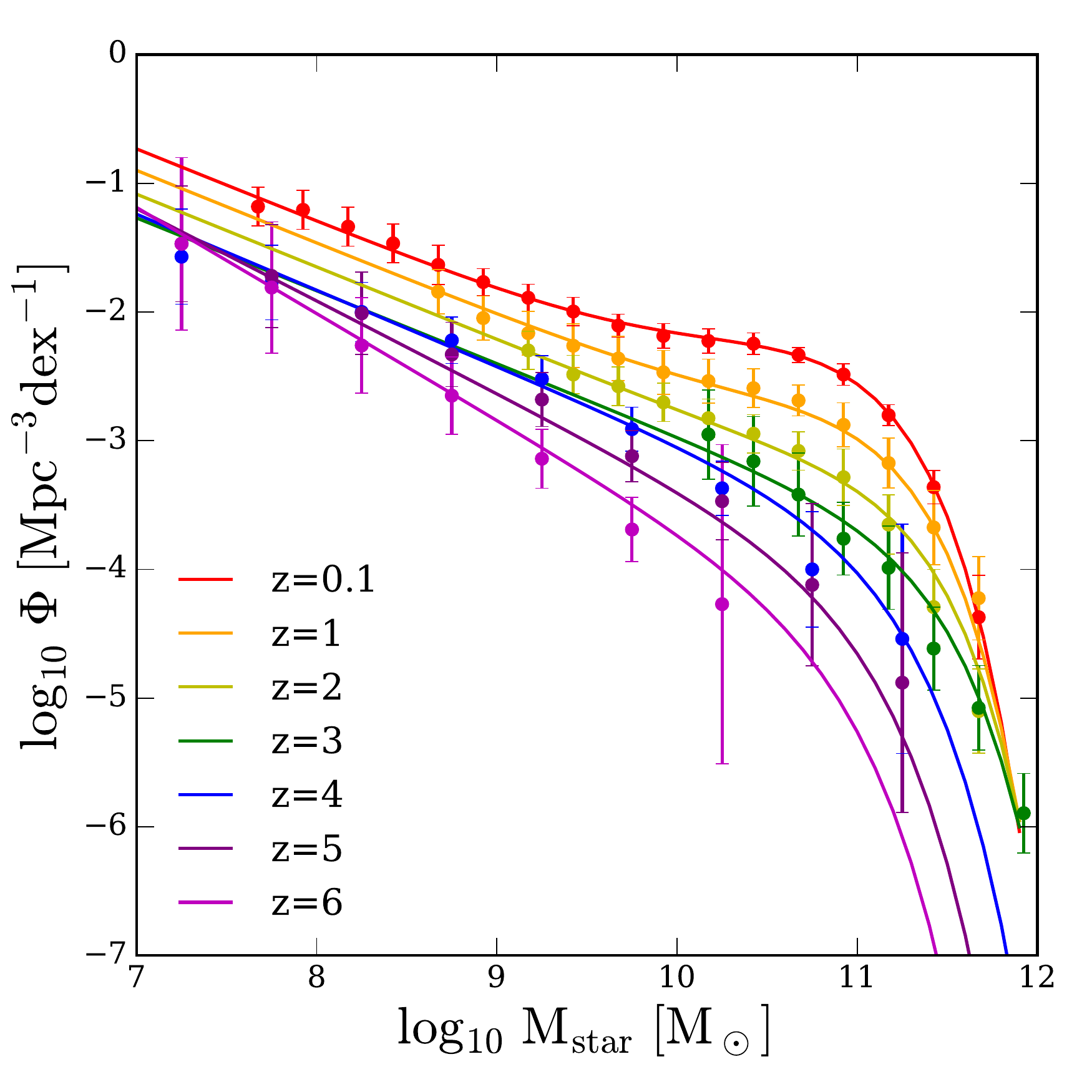}
\includegraphics[width=0.67\columnwidth]{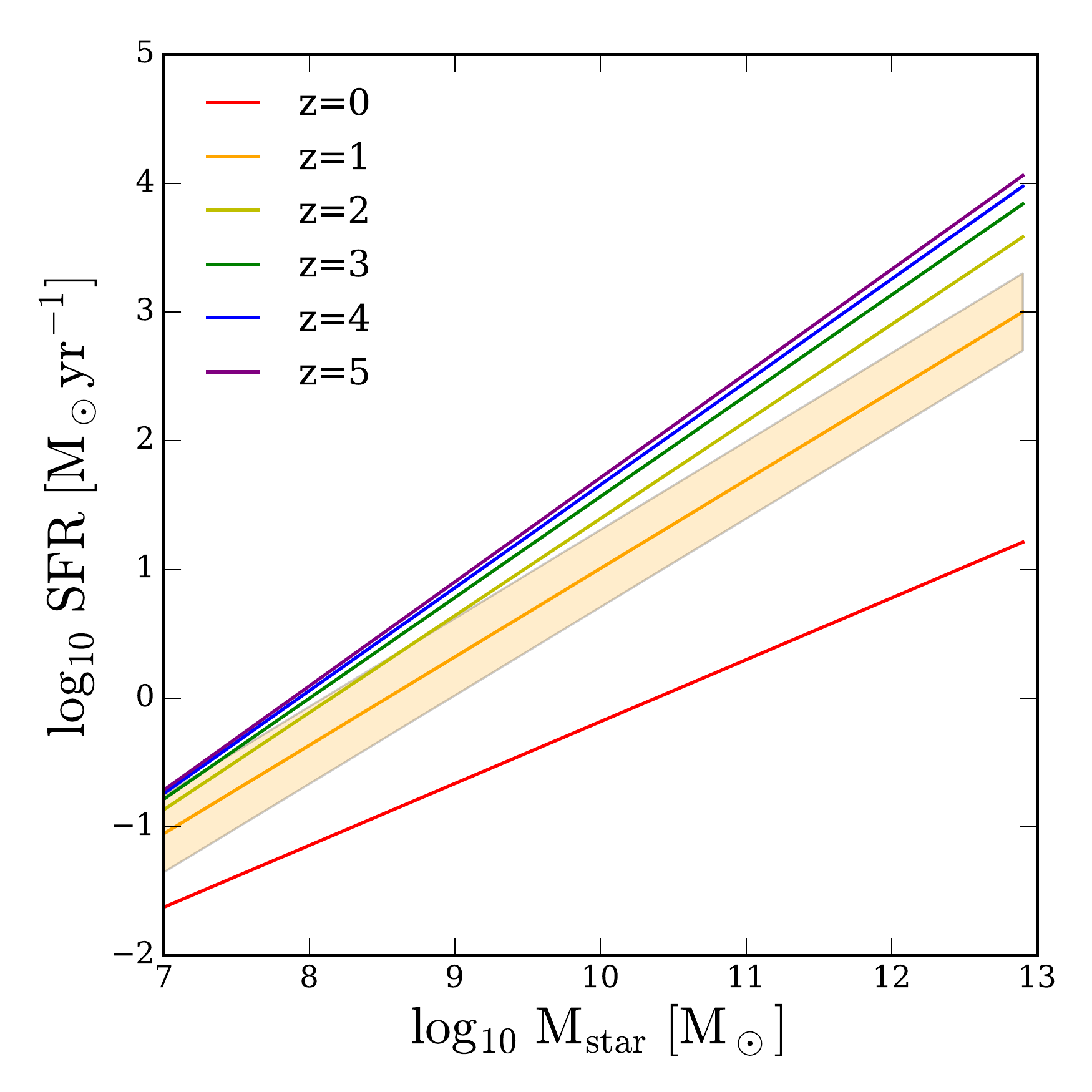}
\includegraphics[width=0.67\columnwidth]{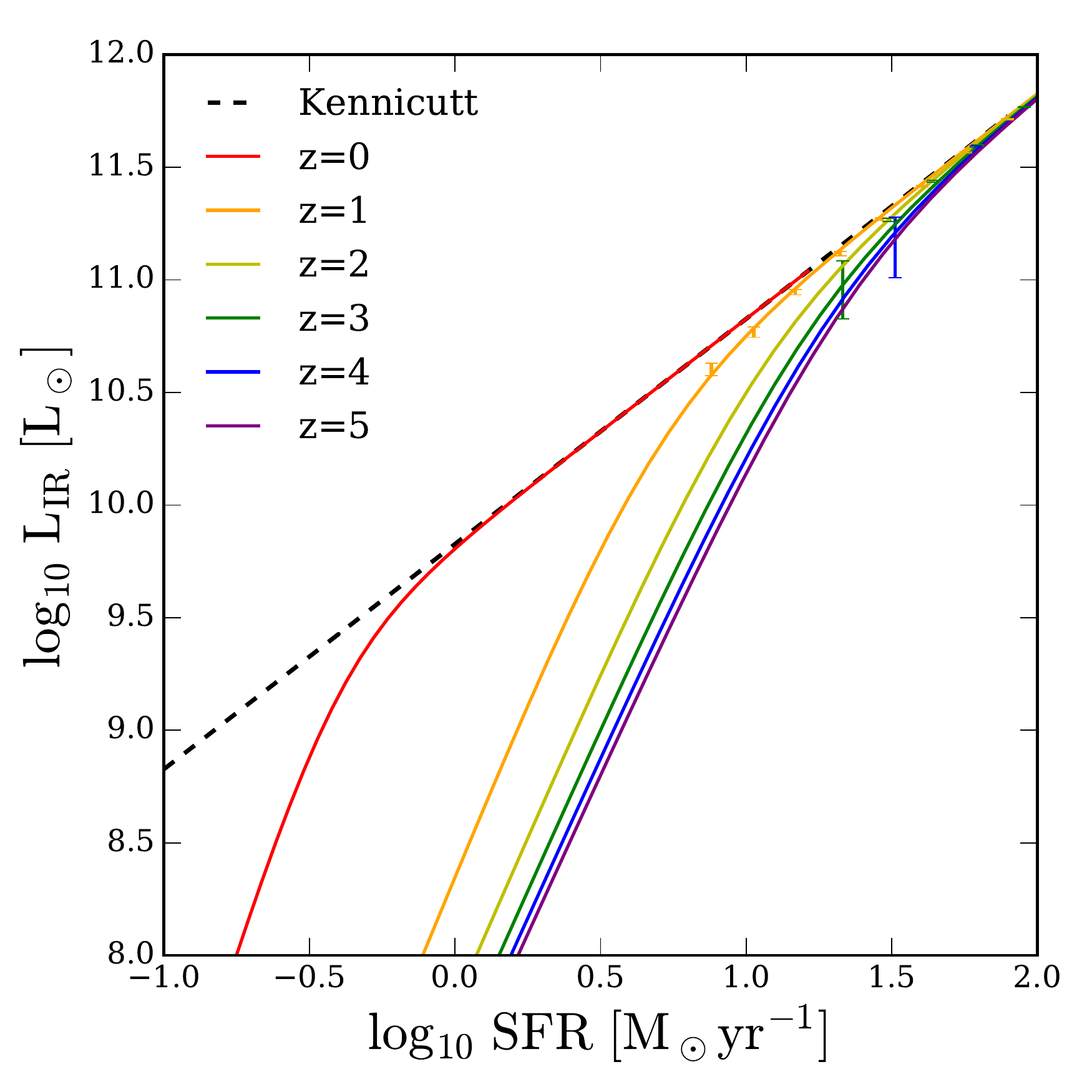}
\caption[]{Key elements of our model. {Left-hand panel}: stellar mass functions from \protect\cite{Henriques15} and \protect\cite{Song16}.  We fit redshift-dependent Schechter functions to the data and preform abundance matching between  $\Ms$ and $\vpk$ (see Section~\ref{sec:abmatch} and Appendix~\ref{app:smf}). {Centre}: star-forming main sequence from \protect\cite{Speagle14}, which is used to assign $\SFR$ to $\Ms$ (see Section~\ref{sec:SFR_Ms}).  {Right-hand panel}: $\LIR$--$\SFR$ relation based on the IRX--$\Ms$ relation from \protect\cite{Heinis14}.  The dashed line corresponds to $\LIR\propto\SFR$ (the Kennicutt relation), which predicts too high $\LIR$ for low-mass galaxies (see Section~\ref{sec:LIR_SFR}).}
\label{fig:model}
\end{figure*}

Steps (ii), (iii), and (iv) are demonstrated in Figure~\ref{fig:model}. 
Below we describe each step in detail.

\subsection{Dark matter haloes from the Bolshoi--Planck simulation}\label{sec:BolshoiP}
We use the public halo catalogues of the Bolshoi--Planck simulation \citep{Klypin16BolshoiP,Rodriguez-Puebla16}\footnote{\href{http://hipacc.ucsc.edu/Bolshoi/MergerTrees.html}{http://hipacc.ucsc.edu/Bolshoi/MergerTrees.html}}, which is based on a Lambda cold dark matter cosmology consistent with the {\em Planck} 2013 results \citep{Planck13cosmo}:
$\Omega_\Lambda$ = 0.693;
$\Omega_{\rm M}$ = 0.307;
$\Omega_{\rm b}$ = 0.048;
$h$ = 0.678;
$n_{\rm s}$ = 0.96;
and $\sigma_8$ = 0.823.
The simulation has a box size of 250 $h^{-1}$Mpc and 
a mass resolution of $1.5\times10^{8} h^{-1}\Msun$. 

The simulation is processed with {\sc rockstar} halo finder \citep{Behroozi13rs} and {\sc consistent trees} \citep{Behroozi13tree}.  Therefore, the halo catalogues include the mapping between central haloes and subhaloes, as well as the peak circular velocity of a halo in its history ($\vpk$).  In this work, we use all haloes with $\vpk >$ 100 km s$^{-1}$ between $z=0.25$ and $5$, with a redshift interval of $\Delta z \approx 0.25$.  When calculating theoretical uncertainties (see Section~\ref{sec:results}), we use 0.1\% of the haloes in the simulation ($\sim$ 6000 haloes in the $z=0.25$ snapshot) to facilitate the calculation.

\subsection{Stellar mass from abundance matching}\label{sec:abmatch}

To assign a stellar mass to each halo, we perform abundance matching between $\vpk$ and observed stellar mass functions.  The basic concept of abundance matching is to assign higher stellar masses to more massive haloes based on the number density, either monotonically or with some scatter \citep[e.g.,][]{ValeOstriker04,Shankar06,Behroozi13,Moster13}.  Instead of halo mass, we use $\vpk$, which is less affected by mass stripping and better correlated with stellar mass \citep[e.g.,][]{NagaiKravtsov05, Conroy06, Wang06, WetzelWhite10,Reddick13}.

First, we collect observed stellar mass functions from the literature. For $z\leq3$, we use the recent compilation of stellar mass functions by \citet[][see their figures 2 and A1]{Henriques15}\footnote{The data sets are publicly available at \href{http://galformod.mpa-garching.mpg.de/public/LGalaxies/figures_and_data.php}{http://galformod.mpa-garching.mpg.de/public/LGalaxies/figures$\_$and$\_$data.php}.}, which are calibrated with the {\em Planck} cosmology. Following \cite{Henriques15}, we add $\Delta \Ms= 0.14$ to convert to the BC03 SPS model. For $z \ge 4$, we use the stelar mass functions  by \citet[][see their table 2]{Song16}, which are derived from the rest-frame UV observations from CANDELS, GOODS, and HUDF, based on the BC03 SPS model.

Secondly, we fit the stellar mass functions using redshift-dependent Schechter functions (see Appendix~\ref{app:smf}).  For $ 0 \leq z\leq 3.5$, we use a double Schechter function with constant faint-end slopes; for $ 3.5 < z \leq 6$, we use a single Schechter function with a time-dependent slope. Using the fitting functions presented in Appendix~\ref{app:smf}, we are able to interpolate smoothly between redshifts. The left-hand panel of Figure~\ref{fig:model} shows the data points and the fitting functions.  Although we fit the stellar mass function out to $z=6$, we only use galaxies at $z\leq5$ in our calculations.

Thirdly, we perform abundance matching between the stellar mass functions and the $\vpk$ of haloes, assuming a scatter of 0.2 dex \citep[e.g.,][]{Reddick13}. In the calculation, the input stellar mass function is first deconvolved with the scatter, and then the deconvolved stellar mass function is used to assign $\Ms$ to $\vpk$ monotonically. We use the code provided by Y.-Y. Mao\footnote{\href{https://bitbucket.org/yymao/abundancematching}{https://bitbucket.org/yymao/abundancematching}}, which follows the implementation in \cite{Behroozi10,Behroozi13}.  With this step, a stellar mass is assigned to each halo.

\subsection{SFR from the star-forming main sequence}\label{sec:SFR_Ms}

We assign an SFR to each $\Ms$ based on the star-forming main sequence compiled by \cite{Speagle14}:
\beqa
\log_{10} \SFR(\Ms, t) =& (0.84 - 0.026 \times t) \log_{10} \Ms \\ &
- (6.51- 0.11 \times t) \ ,
\eeqa 
where $t$ is the age of the universe in Gyr.  This relation is shown in the central panel of Figure~\ref{fig:model}.  The compilation of \cite{Speagle14} is based on the Kroupa IMF, the BC03 SPS model, and the cosmological parameters $\Omega_\Lambda$ = 0.7, $\Omega_{\rm M}$ = 0.3, and $h$ = 0.7.  This cosmology is slightly different from our choice; however, these authors stated that the effect of cosmology is negligible for the main-sequence calibration.

In our calculation, for each $\log_{10}\Ms$, an SFR is drawn from a normal distribution with a mean given by the equation above and a scatter of 0.3 dex.  We note that \cite{Speagle14} have shown that the intrinsic scatter (deconvolved with the evolution in a redshift bin) and the true scatter (excluding observational uncertainties) of the main sequence are 0.3 and 0.2 dex, respectively.  We find that a scatter of 0.2 dex produces too low number counts and too low shot noise (see Section~\ref{sec:results}).  In the central panel of Figure~\ref{fig:model}, we show a 0.3 dex of scatter around the mean relation at $z=1$.

\subsection{Infrared luminosity from SFR and stellar mass}\label{sec:LIR_SFR}

To calculate $\LIR$, it is commonly assumed that $\LIR \propto \SFR$ \citep[the Kennicutt relation; ][]{Kennicutt98,KennicuttEvans12}.  However, this relation is known to break down for low-mass galaxies, which tend to have lower dust content, lower attenuation, and lower $\LIR$ \citep[e.g.,][]{Pannella09,GarnBest10,Buat12,Hayward14}.  One way to improve upon the Kennicutt relation is to assume that the photons produced by star formation are split into UV and IR,
\beq
\SFR = \KUV \LUV + \KIR \LIR \ ,
\eeq
and then use a relation between $\LIR$ and $\LUV$  \citep[e.g.,][]{Bernhard14}.  The logarithm of the ratio between $\LIR$ and $\LUV$ is commonly referred to as the IR-excess (IRX),
\beq
\IRX = \log_{10}\left(\frac{\LIR}{\LUV}\right)\ ,
\eeq
and has been calibrated observationally.  Given the two equations above, we can solve for $\LIR$:
\beq
\LIR = \frac{\SFR}{\KIR + \KUV 10^{-\IRX(\Ms)}} \ .
\eeq
We use $\KUV = 1.71\times10^{-10}$ and $\KIR = 1.49\times10^{-10}$ from \cite{KennicuttEvans12} based on the Kroupa IMF . 

\cite{Heinis14} calibrated the IRX--stellar mass relation based on the rest-frame UV-selected galaxies at $z\sim$ 1.5, 3, and 4 in the COSMOS field observed with {\em Herschel}-SPIRE (part of the HerMES program).  They provided the fitting function 
\beq
\IRX(\Ms) = \alpha \log_{10} \left( \frac{\Ms}{10^{10.35}\Msun} \right) + \IRX_0 \ ,
\eeq
where $\IRX_0 = 1.32$ and $\alpha=0.72$.  We adopt $\alpha=1.5$, which agrees better with the CFIRB amplitudes and is still consistent with their observations (see below). 

The right-hand panel of Figure~\ref{fig:model} demonstrates the $\LIR$--$\SFR$ relation with this $\IRX(\Ms)$, which produces lower $\LIR$ for low-SFR galaxies compared with the Kennicutt relation \citep[][the dashed line]{KennicuttEvans12}.  We show the data points from \cite{Heinis14} to demonstrate the level of uncertainties in observations.  In particular, we use the $\Ms$ and $\IRX$ from their figure 3 and calculate the corresponding $\SFR$ and $\LIR$.  We note that the current observations can only constrain the brightest end, and we need to extrapolate to the faint end.  As we will discuss in Section~\ref{sec:results-CFIRB}, this mass-dependent attenuation is essential for reproducing the observed CFIRB amplitudes.

\subsection{Spectral energy distribution}\label{sec:SED}

With the $\LIR$ calculated above, we need the SED $\Theta_\nu$ to calculate the spectral flux density $S_\nu$.  The spectral luminosity density is given by 
\beq
L_\nu = \LIR \Theta_\nu \ ,
\eeq
and $S_\nu$ at the observed frequency $\nu$ is given by
\beq
S_\nu = \frac{L_{(1+z)\nu}}{4\pi \chi^2(1+z)} \ ,
\eeq
where $\chi$ is the comoving distance, and $L_{(1+z)\nu}$ is evaluated at the rest-frame frequency $(1+z)\nu$.

We assume that the SED of each galaxy is given by a single-temperature modified blackbody, 
\beq
\Theta_\nu \propto \nu^{\beta} B_\nu(T_{\rm d})  \ ,
\eeq
where $B_\nu$ is the Planck function, $T_{\rm d}$ is the dust temperature, and $\beta$ is the spectral index.  The SED is normalized such that $\int {\rm d}\nu\Theta_\nu = 1$. We adopt $\beta = 2.1$ based on our previous work for CFIRB \citep{Wu16}, and we note that $\beta=2$ is widely used and theoretically motivated \citep{DraineLee84,MathisWhiffen89}.

To calculate the $T_{\rm d}$ of each galaxy, we adopt the relation between $T_{\rm d}$ and specific star formation rate (SSFR, $\SFR/\Ms$) given by \cite{Magnelli14},
\beq
T_{\rm d} = 98\ [K] \times (1+z)^{-0.065} +6.9 \log_{10} \SSFR \ ,
\eeq
and we assume a normal distribution with a scatter of 2 K around this relation (consistent with their figure 10).  This relation is derived from galaxies up to $z\sim2$ from the PEP and HerMES programs of {\em Herschel} with multiwavelength observations.  The stellar mass is derived from SED fitting, while the SFR is derived by combining UV and IR.  These authors bin galaxies based on $\SFR$, $\Ms$, and $z$ and calculate $T_{\rm d}$ using the stacked far-infrared flux density in each bin. They have found that the $T_{\rm d}$--SSFR relation is tighter than the $T_{\rm d}$--$\LIR$ relation.

\section{Calculating the CFIRB angular power spectra}\label{sec:obs}

With the prediction of $S_\nu$ for each halo in the catalogues, we proceed to compute the CFIRB angular power spectra. The formalism presented below is motivated by the analytical halo model presented in \cite{Shang12}, and we make various generalization and adjustments for our sampling approach. Since we use subhaloes from an $N$-body simulation, we expect our approach to be more accurate than a purely analytical calculation.

The CFIRB auto angular power spectrum is given by the sum of the two-halo term, the one-halo term, and the shot noise:
\beq
C^{\nu}_\ell = C^{\nu, \rm 2h}_\ell + C^{\nu, \rm 1h}_\ell  + C^{\nu, \rm shot}_\ell  \ .
\eeq
Here we present the equations for a single frequency; the equations for two-frequency cross-spectra can be generalized easily.

The two-halo term corresponds to the contribution from two galaxies in distinct haloes and is given by
\beq
C^{\nu, \rm 2h}_\ell = \int \chi^2 {\rm d}\chi F^2_\nu(z) P_{\rm lin}\left(k=\frac{\ell}{\chi}, z\right)  \ ,
\eeq
where $P_{\rm lin}(k,z)$ is the linear matter power spectrum calculated with CAMB \citep{Lewis00}, and 
\beq
\label{eq:K_nu_z}
F_\nu(z) = \int {\rm d}M \frac{{\rm d}n}{{\rm d}M} b(M) \left( S_\nu^{\rm cen} + \int {\rm d}M_{\rm s} \frac{{\rm d}N(M)}{{\rm d}M_{\rm s}} S_\nu^{\rm sat} \right) \ ,
\eeq
where $M$ is the mass of central haloes, ${\rm d}n/{\rm d}M$ and $b(M)$ are the mass function and halo bias of central haloes, $M_s$ is the mass of subhaloes, and ${\rm d}N(M)/{\rm d}M_{\rm s}$ is the number of subhaloes in a central halo. In our sampling approach, the integration is replaced by the sum over all $b(M) S_\nu$, and for a satellite galaxy we use the $b(M)$ of its central halo.  For $b(M)$, we use the fitting function of halo bias from \cite{Tinker10}, and we have verified that this fitting function agrees with the linear halo bias measured directly from the Bolshoi--Planck simulation.

The one-halo term corresponds to the contribution from two galaxies in the same halo and is given by
\beq
C^{\nu, \rm 1h}_\ell = \int \chi^2 {\rm d}\chi G_\nu\left(k=\ell/\chi,z\right) \ ,
\eeq
where
\beqa
G_\nu(k, z) =& 2 \int {\rm d}M\frac{{\rm d}n}{{\rm d}M} S_\nu^{\rm cen} \left( \int {\rm d}M_{\rm s} \frac{{\rm d}N(M)}{{\rm d}M_{\rm s}} S_\nu^{\rm sat} \right)u(k, z) \\
&+ \int {\rm d}M\frac{{\rm d}n}{{\rm d}M} \left( \int {\rm d}M_{\rm s} \frac{{\rm d}N(M)}{{\rm d}M_{\rm s}} S_\nu^{\rm sat} \right)^2 u^2(k,z) \ .
\eeqa
Here, $u(k,z)$ is the density profile of dark matter haloes in the Fourier space, and $u(k,z)\approx 1$  for the large scales discussed in this work. The first term corresponds to summing over the central--satellite pairs in a halo, and the second term corresponds to summing over the satellite--satellite pairs in a halo. We avoid self-pairs in calculating the second term.

The shot noise corresponds to self-pairs of galaxies and is given by
\beq
C^{\nu, \rm shot}_\ell = \int \chi^2 {\rm d}\chi  \int {\rm d}S_\nu \frac{{\rm d}n}{{\rm d}S_\nu} S_\nu^2  \ ,
\eeq
where $S_\nu$ includes both central and satellite galaxies.

The cross angular spectrum between CFIRB and CMB lensing potential is given by
\beqa
C^{\phi\nu}_{\ell} = & \int_0^{\chi_*} \chi^2 {\rm d}\chi(1+z) F_\nu(z) 
\frac{3}{\ell^2}\Omega_{\rm M} H_0^2 \left(\frac{\chi_*-\chi}{\chi_* \chi}\right) \\
&\times P_{\rm lin}\left(k=\frac{\ell}{\chi}, z\right) \ ,
\eeqa
where $\chi_*$ is the comoving distance to the last-scattering surface, and $F_\nu(z)$ is given by Equation~\ref{eq:K_nu_z}.

\section{Comparison with observations}\label{sec:results}
\begin{figure*}
\vspace{-0.5cm}
\centerline{\includegraphics[width=\columnwidth]{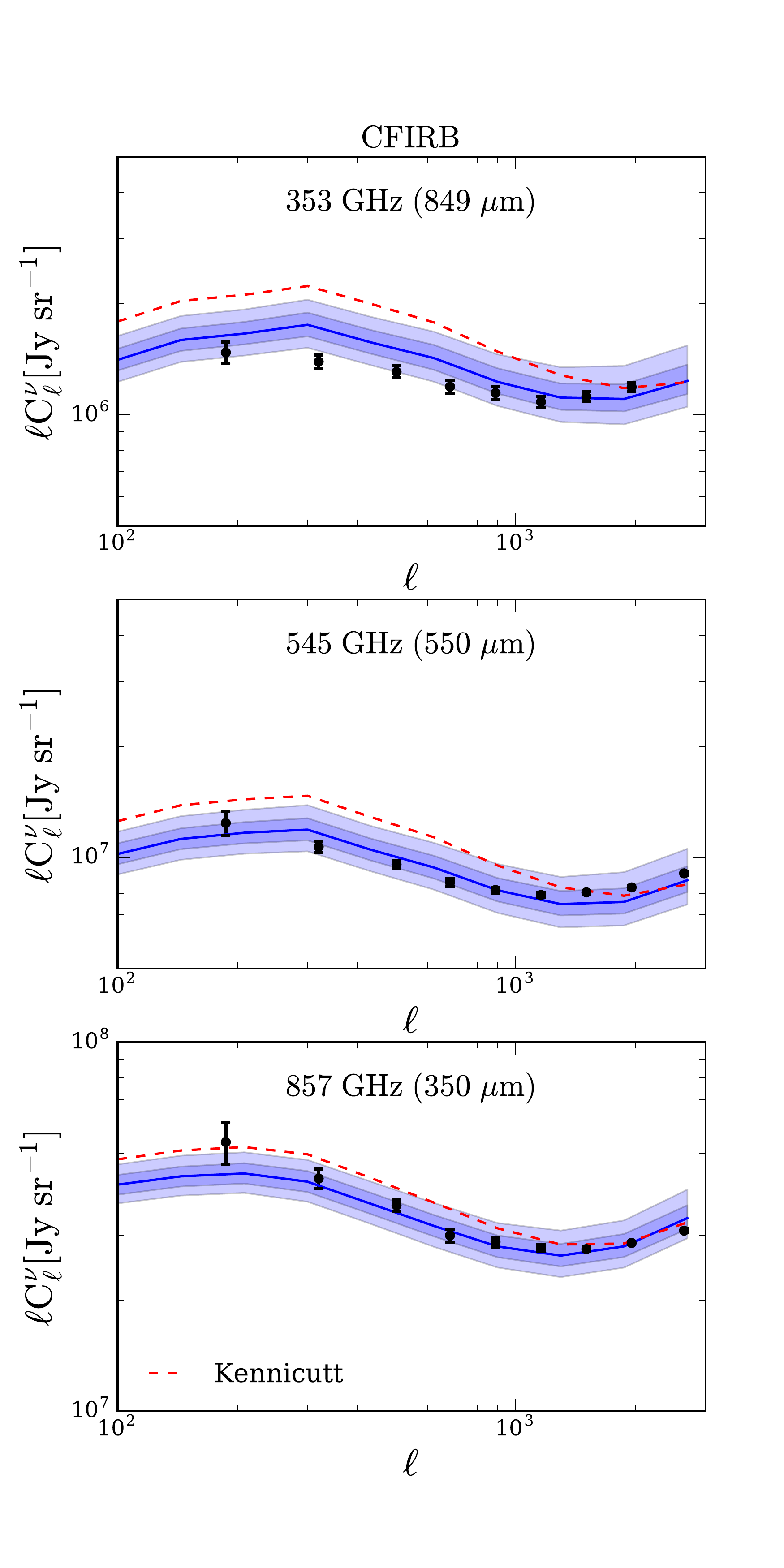}
\includegraphics[width=\columnwidth]{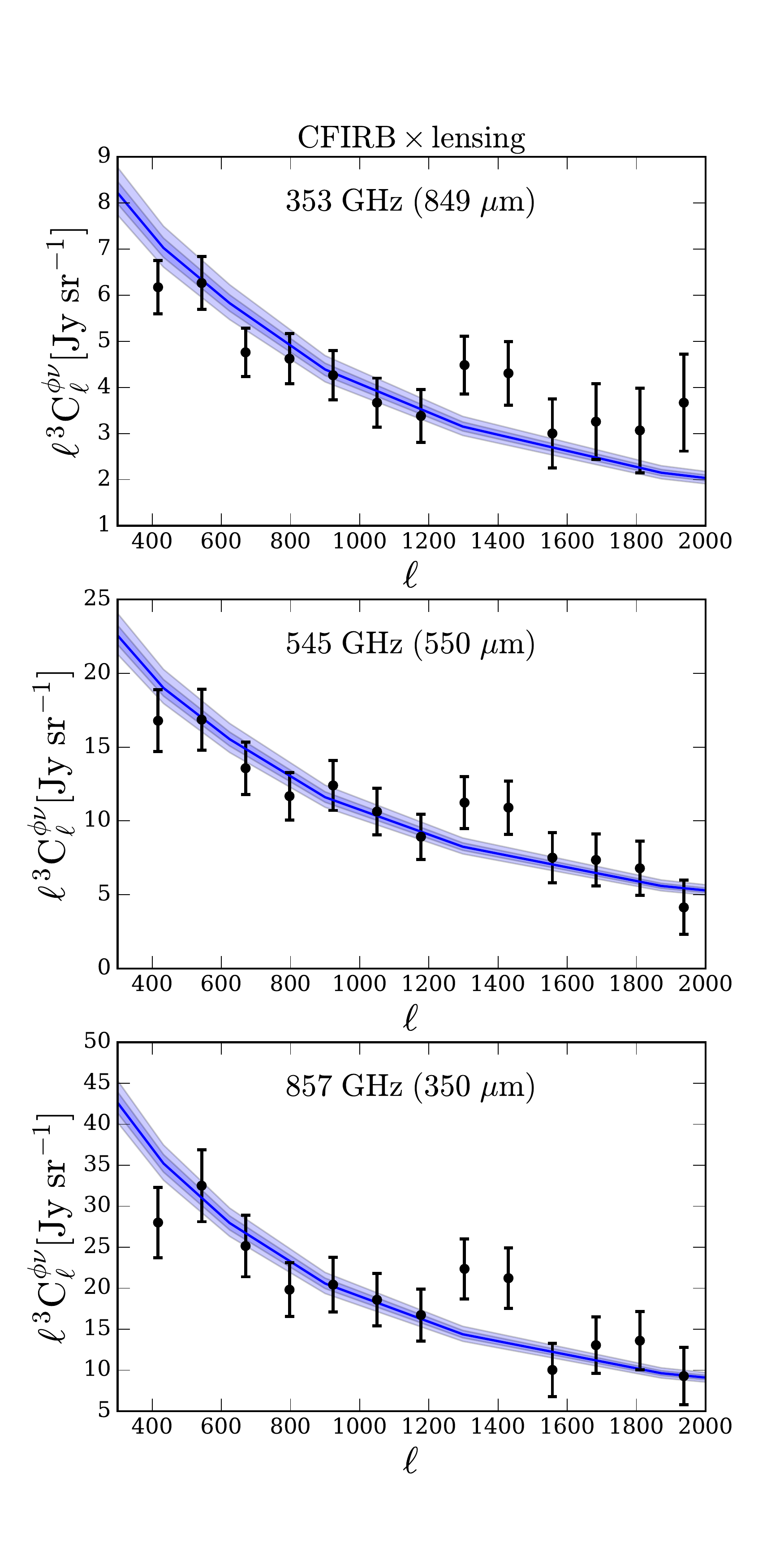}}
\vspace{-1cm}
\caption[]{Comparison between our model (blue bands) and the CFIRB anisotropies observed by {\em Planck} (data points). {Left-hand panel}: CFIRB auto angular power spectra from \cite{Planck13XXX}.  The red dashed curves show that the Kennicutt relation overproduces the large-scale amplitudes.  {Right-hand panel}: cross-angular power spectra between CFIRB and CMB lensing potential from \cite{Planck13XVIII}.  The dark and light blue bands correspond to the 68\% and 95\% intervals of theoretical uncertainties, respectively.}
\label{fig:CFIRB}
\end{figure*}

In this section we compare our model predictions with observational results.

\subsection{CFIRB anisotropies}\label{sec:results-CFIRB}

We compare our model with the CFIRB anisotropies observed by {\em Planck}:
\begin{itemize}
\item \cite{Planck13XXX} presents the CFIRB observed by {\em Planck}-HFI
for an area of 2240 deg$^2$, for which {\sc HI} maps are available for removing the foreground Galactic dust emission. The primordial CMB, the Sunyaev--Zeldovich effect, and the radio sources are also removed. We compare our model with the CFIRB angular power spectra for $187 \leq \ell \leq 2649$, presented in their table D.2.

\item \cite{Planck13XVIII} presents the first detection of the cross-correlation between CFIRB and CMB lensing potential (the latter is extracted from the low-frequency bands of {\em Planck}).  The CMB lensing potential is dominated by dark matter haloes between $z\approx1$ and $3$, and CFIRB is dominated by galaxies in the same redshift range; therefore, the cross-correlation between CFIRB and CMB lensing potential directly probes the connection between FIR galaxies and dark matter haloes.  In addition, compared with the auto-correlation of CFIRB, this cross-correlation is less affected by the contamination of Galactic dust. 
\end{itemize}

Figure~\ref{fig:CFIRB} compares our model predictions with the observational results described above. We include the results in 353, 545, and 857 GHz (849, 550, and 350 $\micron$), and we exclude 217 GHz because CMB dominates this band for all angular scales.  In all calculations, we apply the colour-correction factors and flux cuts of \citet[][see their section 5.3 and table 1]{Planck13XXX}.   The left column corresponds to the auto angular power spectra of CFIRB, $C^{\nu}_\ell$, while the right column corresponds to the cross-angular spectra between CFIRB and CMB lensing potential, $C^{\phi\nu}_\ell$.  To calculate the theoretical uncertainties, we repeat Steps (iii) to (v) in Section~\ref{sec:model} for 1000 times,  and we use 0.1\% of the haloes in the Bolshoi--Planck simulation to facilitate the calculation.  The dark and light blue bands correspond to the 68\% and 95\% intervals of the theoretical uncertainties. As can be seen, our model well captures both observational results.  We emphasize that we perform no fitting to the data, and that all the components of our model directly come from independent surveys of UV, optical, and FIR.

The red dashed curves in the left column of Figure~\ref{fig:CFIRB} show that, if we assume the Kennicutt relation ($\LIR\propto\SFR$) instead of the mass-dependent dust attenuation, we produce too high large-scale amplitudes of the power spectra.   The Kennicutt relation assigns too high $\LIR$ to low-mass galaxies, and because of the high number density of low-mass galaxies, it leads to too high CFIRB large-scale amplitudes.  In this sense, CFIRB can be used to constrain the SFR and dust content of low-mass galaxies.  We note that the Kennicutt relation and the mass-dependent attenuation produce very similar small-scale auto power spectra.  The reason is that the two models have very similar $\LIR$ for massive galaxies, and the small-scale spectra are dominated by shot noise, which is contributed mostly by massive galaxies.

\subsection{Number counts}\label{sec:results-NC}

In this section, we turn to submm number counts, which are dominated by massive galaxies.  We compare our model with the number counts observed by {\em Herschel}-SPIRE at 250, 350, and 500 $\micron$ (1200, 857, and 600 GHz):

\begin{itemize}

\item \cite{Bethermin12} presented the deep number counts from the HerMES survey (the COSMOS and GOODS-N fields). They performed stacked analyses based on the 24 $\micron$ sources, and they managed to extract the number counts down to $\sim$ 2 mJy.

\item \cite{Valiante16} presented the number counts from the H-ATLAS survey of an area of 161.6 deg$^2$ (the GAMA fields). At the faint end, their results agree with \cite{Bethermin12}; at the bright end, they have better statistics due to the larger survey area. 

\end{itemize}

\begin{figure}
\vspace{-0.5cm}
\includegraphics[width=\columnwidth]{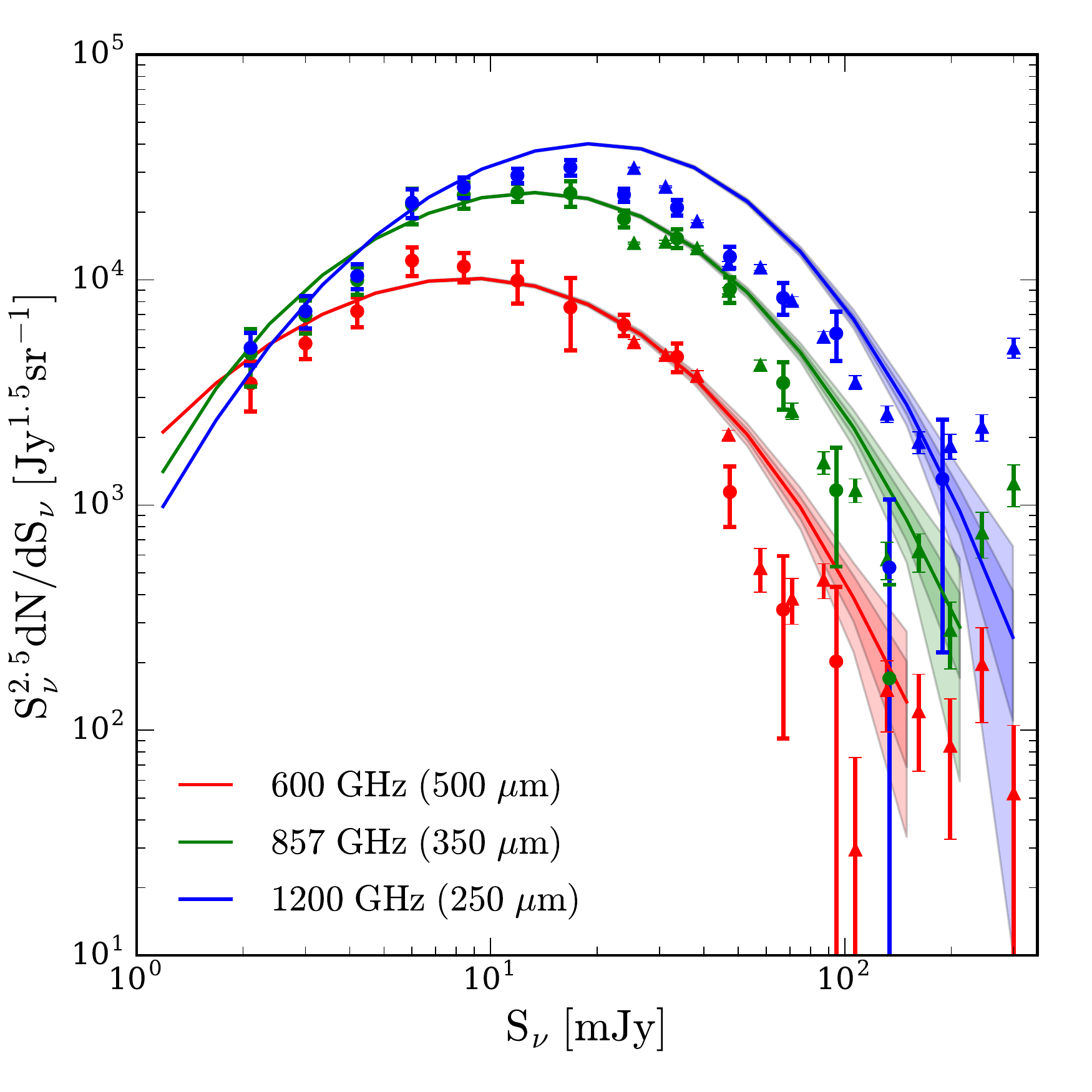}
\vspace{-0.5cm}
\caption[]{Number counts predicted from our model (colour bands) compared with the results from HerMES \protect\citep[][circles]{Bethermin12} and H-ATLAS \protect\citep[][triangles]{Valiante16}. The dark and light bands correspond to the 68\% and 95\% intervals of the theoretical uncertainties, respectively.}
\label{fig:NC}
\end{figure}

Figure~\ref{fig:NC} compares the number counts from our model with the two observational results described above.  For this calculation, we use all haloes in the Bolshoi--Planck simulation to obtain enough bright galaxies. We note that the observed brightest end ($\gtrsim100$ mJy) is dominated by gravitationally lensed sources, which we do not have in our model.  Our model mostly agrees with the observational results; however, for 250 $\micron$ (1200 GHz) our model produces slightly higher number counts; this could result from our oversimplified assumption for SED.

\section{Discussions}\label{sec:discussions}

In our model, we assume that all galaxies belong to the star-forming main sequence.  This is a simplified assumption, because it is known that a fraction of massive galaxies are quiescent \citep[e.g.,][]{Ilbert13,Moustakas13,Muzzin13,Tomczak14,Man16,Schreiber16}.  In addition, studies have also shown that quiescent galaxies can still have significant FIR emission due to the dust heated by old stars (the so-called cirrus dust emission, e.g., \citealt{Fumagalli14,Hayward14,Narayanan15}).  In \cite{Wu16}, we have also found  that the observed CFIRB requires substantial FIR emission from massive haloes.

We have attempted to include quiescent galaxies in this work, but we find that the CFIRB data cannot distinguish between $\LIR$ coming from star formation and cirrus dust.  When we include a fraction of quiescent galaxies with SSFR = $10^{-12} \rm yr^{-1}$ \citep[e.g.,][]{Muzzin13, Fumagalli14}, the power spectra are lowered, and we need to add cirrus dust emission to these quiescent galaxies to compensate for the lowered power.   However, the fraction of quiescent galaxies and the cirrus dust emission are both highly uncertain are degenerate with each other; therefore, we decide not to include them in this work. Investigating the contribution from quiescent galaxies will require the modelling of old stars and cirrus dust, as well as comparisons with near-infrared observations.  We will investigate this in future work.

Furthermore, it is also known that a small fraction of galaxies undergo starburst phases and have significantly higher SFR and IR luminosities \citep[e.g.,][]{Elbaz11}.  The starburst galaxies account for $\sim10\%$ of the cosmic SFR density at $z\sim2$  \citep{Rodighiero11,Sargent12}  and are expected to have negligible contribution to the CFIRB \citep[e.g.,][]{Shang12,Bethermin13}.  The effect of starburst galaxies will be degenerate with that of quiescent galaxies in producing CFIRB.  Therefore, any departure from the star-forming main sequence will require constraints from multiwavelength observations, which will be explored in our future work.

In this work, we choose a minimal number of modelling steps in order to avoid degeneracies.  Except for the slope of the IRX--$\Ms$ relation, all the other parameter values are directly taken from the literature, and we neither introduce new parameters nor fit parameters to the data.  In our future work, we plan to incorporate more astrophysical processes (including quiescent and starburst galaxies, realistic SEDs) into our model,  combine multiwavelength observational results from UV, optical, near-IR, FIR, and radio surveys, and perform Markov chain Monte Carlo calculations to constrain model parameters, 

Over the next decade, new instruments are expected to revolutionize the view of the FIR/submm sky.   The Far-infrared Surveyor (Origins Space Telescope), which is currently planned by NASA, prioritizes the measurements of cosmic SFR.  The Cosmic Origins Explorer  (CORE, \citealt{CORE16a}) and the ground-based CMB-S4 experiment \citep{Abazajian16} will measure CFIRB and CMB lensing to unprecedented precision.  The Primordial Inflation Explorer (PIXIE, \citealt{Kogut11}) will significantly improve the accuracy of the absolute intensity of CFIRB compared with {\em COBE}-FIRAS. These missions are expected to lead to a consistent picture of cosmic star formation history.  In a companion paper, we apply a principle component approach to investigate the optimal experimental designs for constraining the cosmic star-formation history using CFIRB \citep{Wu16c}.  We plan to apply the empirical approach presented in this paper to generate mock catalogues, check consistencies between models, and develop survey strategies for these observational programs. 

\section{Summary}\label{sec:summary}

We present a minimal empirical model for dusty star-forming galaxies to interpret the observations of CFIRB anisotropies and submm number counts.  Our model is based on the Bolshoi--Planck simulation and various results from UV/optical/IR galaxy surveys.  Below we summarize our model and findings:

\begin{itemize}

\item To assign IR spectral flux densities $S_\nu$ to dark matter haloes, we model stellar mass (using abundance matching between $\vpk$ and observed stellar mass functions), SFR (using the star-forming main sequence), $\LIR$ (assuming a mass-dependent attenuation), and SED (assuming a modified blackbody).

\item Given the connection between $S_\nu$ and halo mass obtained above, we apply an extended halo model to calculate the auto angular power spectra of CFIRB and the cross-angular power spectra between CFIRB and CMB lensing potential. We find that the commonly used Kennicutt relation, $\LIR \propto\SFR$, leads to too high CFIRB amplitudes. The observed CFIRB amplitudes require that low-mass galaxies have lower $\LIR$ than expected from the Kennicutt relation.  This trend has been observed previously and is related to the low dust content of low-mass galaxies.

\item Our model also produces submm number counts that agree with observational results of {\em Herschel}.  The number counts are contributed by massive haloes, and this agreement indicates that our minimal model (star-forming main sequence only, no quiescent or starburst galaxies) is sufficient for dusty star-forming galaxies in massive haloes.  We slightly overproduce the number counts at 250 $\micron$ (1200 GHz), and this may indicate that the SEDs of dust emission deviate from a simple modified blackbody.
 
\end{itemize}

Our results indicate that the observed CFIRB broadly agrees with the current knowledge of galaxy evolution from resolved galaxies in UV and optical surveys, under the assumption that low-mass galaxies produces IR luminosities lower than expected from the Kennicutt relation.   Therefore, CFIRB provides a rare opportunity of constraining the SFR and dust production in low-mass galaxies.  However, since CFIRB does not provide redshifts of galaxies, further investigations for low-mass galaxies will require the cross-correlation between CFIRB with galaxies or extragalactic background light observed in other wavelengths \citep[e.g.,][]{Cooray16,Serra16}.

\section*{Acknowledgements}
We thank Joanne Cohn and Martin White for helpful discussions, and we thank Yao-Yuan Mao for providing the code and assistance for the abundance matching calculation. HW\ acknowledges the support by the US\ National Science Foundation (NSF) grant AST1313037.  The calculations in this work were performed on the Caltech computer cluster Zwicky, which is supported by NSF MRI-R2 award number PHY-096029.  OD\ acknowledges the hospitality of the Aspen Center for Physics, which is supported by NSF grant PHY-1066293.  Part of the research described in this paper was carried out at the Jet Propulsion Laboratory, California Institute of Technology, under a contract with the National Aeronautics and Space Administration.  The Bolshoi--Planck simulation was performed by Anatoly Klypin within the Bolshoi project of the University of California High-Performance AstroComputing Center (UC-HiPACC) and was run on the Pleiades supercomputer at the NASA Ames Research Center.
\bibliographystyle{mnras}
\bibliography{/Users/hao-yiwu/Dropbox/master_refs}

\appendix
\section{Fitting the observed stellar mass functions}\label{app:smf}
In Section~\ref{sec:abmatch},  we fit redshift-dependent Schechter functions to the observed stellar mass functions.  We minimize 
\beq
\chi^2 = \frac{\sum(\log_{10}\Phi_{\rm model} - \log_{10}\Phi_{\rm data})^2}{(\Delta\log_{10}\Phi_{\rm data})^2} \ .
\eeq
For $ 0 \leq z\leq 3.5$, we use a double Schechter function:
\beqa
\Phi(\Ms) &= \frac{{\rm d}n}{{\rm d}\log_{10}\Ms} [{\rm Mpc^{-3}dex^{-1}}] \\
&= (\Phi_1  m^{1+\alpha_1} + \Phi_2 m^{1+\alpha_2} ) e^{-m} \ln(10)  \ ,
\eeqa
where 
\beq
m = \frac{\Ms}{M_0}  \ .
\eeq
We assume that $M_0$, $\Phi_1$, and $\Phi_2$ depend on $z$, while $\alpha_1$ and $\alpha_2$ are independent of $z$. The best-fitting parameters are 
\beqa
\log_{10}M_0 &=10.90 + 0.08 \times z \\
\log_{10}\Phi_1 &= -2.4 -0.61 \times z \\
\log_{10}\Phi_2 &= -3.29 -0.23 \times z \\
\alpha_1 &= -0.68 \\
\alpha_2 &= -1.57 \\
\eeqa
with $\chi^2$ = 7.4 with 41 degrees of freedom.

For $3.5 < z \leq 6$, we use a single Schechter function:
\beq
\Phi(\Ms) =\Phi_1  m^{1+\alpha_1}  e^{-m} \ln(10)  \ ,
\eeq
and we assume that all parameters depend on $\ln(1+z)$.  The best-fitting parameters are 
\beqa
\log_{10}M_0 &= 12.26 -0.77  \times \ln(1+z) \\
\log_{10}\Phi_1 &=-0.77 -1.99 \times \ln(1+z) \\
\alpha_1 &= -0.47 -0.69 \times \ln(1+z) \\
\eeqa
with $\chi^2$ = 3.7 with 19 degrees of freedom.  The fitting functions are shown in the left-hand panel of Figure~\ref{fig:model}.

\bsp	
\label{lastpage}
\end{document}